\documentstyle[prl,aps,multicol]{revtex}
\input psfig
\begin{document}

\draft

\title{Interface Roughening in a Hydrodynamic Lattice-Gas Model
\\ with Surfactant}

\author{Francis~W. Starr, Stephen~T. Harrington, Bruce~M. Boghosian, and
H.~Eugene Stanley}

\address{Center for Polymer Studies, Center for Computational Science,
and Department of Physics, Boston University, Boston, MA 02215 USA}

\date{LT5906: Received 18 June 1996}

\maketitle

\vspace*{-4.5cm}{\flushleft{\footnotesize BU-CCS-960601 }}\vspace*{+4.5cm}

\begin{abstract}

Using a hydrodynamic lattice-gas model, we study interface growth in a
binary fluid with various concentrations of surfactant.  We find that
the interface is smoothed by small concentrations of surfactant, while
microemulsion droplets form for large surfactant concentrations.  To
assist in determining the stability limits of the interface, we
calculate the change in the roughness and growth exponents $\alpha$ and
$\beta$ as a function of surfactant concentration along the interface.

\end{abstract}


\begin{multicols}{2}

The study of rough interfaces is of important experimental and
theoretical interest and has wide interdisciplinary applications,
including fluid imbibition experiments, flow in porous media and
fluid--fluid displacement \cite{vfk}.  Rough interfaces have been
extensively studied by direct integration of continuum equations and
using simple discrete lattice models.  The development of lattice-gas
models makes it possible to study interface roughening in hydrodynamic
systems.  Lattice-gas models have evolved from simple one-component
Navier-Stokes fluids \cite{FHP}, to two-component immiscible lattice-gas
(ILG) models \cite{RK}, and, most recently, to a model including
amphiphilic particles \cite{BCE}.  Lattice-gas models can reproduce the
dynamics on mesoscopic scales, allowing for the investigation of
non-equilibrium behavior over a much broader range of length and time
scales than is possible with molecular dynamics \cite{Rothman}.  The
simplicity of the collision rules, the exact conservation of mass and
momentum, and the natural underlying kinetic fluctuations suggest that
lattice-gas models are an appropriate choice for studying the scaling
behavior of interfaces in complex hydrodynamic systems.

In this Letter, we study interfacial roughening in the presence of a
surfactant by including amphiphilic particles that tend to reside at the
binary fluid interface \cite{bar}.  Flekk\o y and Rothman recently
studied fluctuating interfaces in an ILG model without surfactant and
described the scaling properties of these interfaces \cite{FR}.  When
surfactant is added to an initially flat interface, we find that the
roughness exponent $\alpha$ and growth exponent $\beta$ decrease for
small concentrations of surfactant \cite{exponents}; however, for larger
concentrations of surfactant, $\beta$ increases while $\alpha$ continues
to decrease.  The continued addition of surfactant causes the interface
to spontaneously break up into a microemulsion phase.  We also find that
the saturated width of the interface has a minimum when the surfactant
concentration along the interface matches the binary fluid concentration
in the bulk.

Far from any interface, the present model approaches the one-component
FHP-II model \cite{FHP,Frisch}, which employs a triangular lattice with
each lattice site occupied by up to seven particles, each of which is
either at rest or has unit velocity directed along one of the six
lattice vectors toward neighboring sites.  No two particles may have the
same velocity at a lattice site.  The rules dictate that a particle
propagates along a lattice direction until a collision occurs.  When no
surfactant is present, the current model reduces to the two-component
ILG model \cite{RK}, which employs two types of particle and assigns
each of the two species a color, ``red'' or ``blue''.  Each collision
must conserve red particles, blue particles, and the total momentum.
Additional collision rules are included that give rise to the aggregate
behavior of an immiscible fluid.  These rules can be described using an
electrostatic analogy \cite{BCE}, where the {\it color flux} ${\bf
J}({\bf x}, t)$ of an outgoing state is the difference between the red
and blue momenta.  The {\it color field} ${\bf E}({\bf x}, t)$ is the
gradient of color between neighboring sites.  At each time step, the
{\it color work} -- defined by the inner product ${\bf J} \cdot {\bf E}$
-- is minimized at each lattice site, which has the effect of inducing
phase separation.

With the addition of surfactant molecules, the ILG becomes a
three-component model having the potential to model amphiphilic systems
\cite{BCE}.  To make the behavior of the surfactant consistent with a
molecule composed of hydrophobic and hydrophilic ends, the surfactant is
represented as a ``color dipole'' with a {\it dipole vector} ${\bf
\sigma}({\bf x}, t)$ and produces a {\it dipolar field} ${\bf P}({\bf
x},t)$ at neighboring sites.  A term proportional to ${\bf \sigma} \cdot
{\bf E}$, representing the interaction of colored particles with
dipoles, is added to the Hamiltonian.  Similarly, we add a term
proportional to ${\bf \sigma} \cdot {\bf P}$ to include the
dipole-dipole interaction.  Minimization of the modified Hamiltonian
favors the surfactant particles lining up along the interface between
red and blue particles and makes it unfavorable for surfactant particles

\newbox\figa
\setbox\figa=\psfig{figure=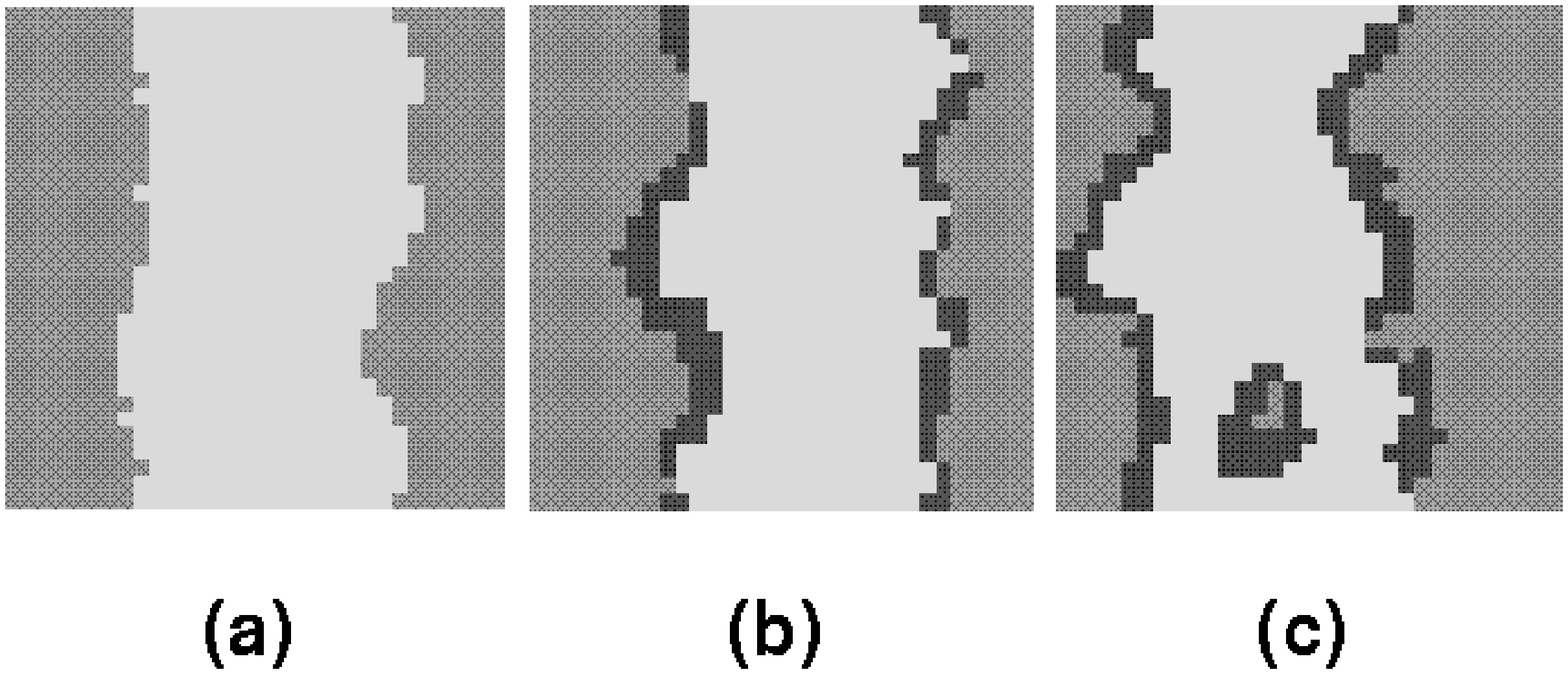,width=3.375in}
\begin{figure*}[htbp]
\begin{center}
\leavevmode
\centerline{\box\figa}
\narrowtext
\caption{ Typical configurations (a) for the $L=32$ case without
surfactant ($\rho_s = 0$); the concentration in each region is $\rho =
0.5$; (b) for a surfactant concentration $\rho_s = 1.0$ along the
interface; binary fluid particles are replaced by surfactant particles
so that the total particle number remains unchanged; (c) for the
micellular microemulsion phase ($\rho_s = 1.5$).  Note that the
periodic boundary conditions require two interfaces.  }
\label{Fig0}
\end{center}
\end{figure*}

\noindent to align with each other at neighboring sites, modeling the expected
behavior of an amphiphilic molecule \cite{coupling}.

We perform simulations of the three-component model using a sequence of
two-dimensional lattices of edge $L=16,32,64,128$ and $256$ with
periodic boundary conditions in both directions.  Initially, red and
blue particles are placed on the lattice separated by two flat
interfaces which are evenly spaced and oriented perpendicular to one of
the lattice directions \cite{adler}, as shown in Fig.~\ref{Fig0}(a).  In
each region the particle concentration is $\rho = 0.5$ \cite{conc}.
Each particle, with uniform probability, is either given zero velocity
or unit velocity in one of the six lattice directions.  Surfactant
molecules are placed between the interfaces with varying concentration
and angular orientation randomly selected between between $0$ and
$2\pi$.  (After a single collision step, the angular orientation of the
surfactant molecules tends to align perpendicular to the interface, as
this is energetically favorable.)  The system then evolves according to
the collision rules of the model.  Our interest focuses on how the
addition of surfactant to the interface mediates the roughening process.

We first calculate the average interface width $W(L,t) \equiv \langle
\overline{h^2({\bf x},t)} - \overline{h({\bf x},t)}^2 \rangle ^{1/2}$ at
logarithmically spaced time intervals with no surfactant present.  We
find that $W(L,t) \sim t^{\beta}$ until a crossover time
$t_{\times}(L)$, whereupon it saturates at a value $W_{sat}(L) \sim
L^{\alpha}$ \cite{vfk,FR}, with $\beta=0.33 \pm 0.02$ and $\alpha = 0.50
\pm 0.02$ (Fig.~\ref{Fig1}).  Hence $z = \alpha / \beta = 1.50 \pm
0.03$, where $z$ is the dynamical exponent, defined by $t_{\times}(L)
\sim L^z$.  In Fig.~\ref{Fig1}(b), we plot the rescaled width
$W(L,t)/L^{\alpha}$ against the rescaled time $t/L^z$ and confirm the
scaling hypothesis, $W(L,t) \sim L^{\alpha}~ f(t/t_{\times})$
\cite{vfk}.  Our results match the scaling exponents $\alpha = 1/2$ and
$\beta = 1/3$ for the Kardar-Parisi-Zhang (KPZ) universality class
\cite{KPZ}, and are comparable to earlier results obtained by Flekk\o y
and Rothman \cite{FR}.

Next we consider interface roughening with surfactant present.  In the
three-component model, for surfactant concentrations below the critical
concentration for spontaneous emulsification, the surfactant particles
tend to

\newbox\figa
\setbox\figa=\psfig{figure=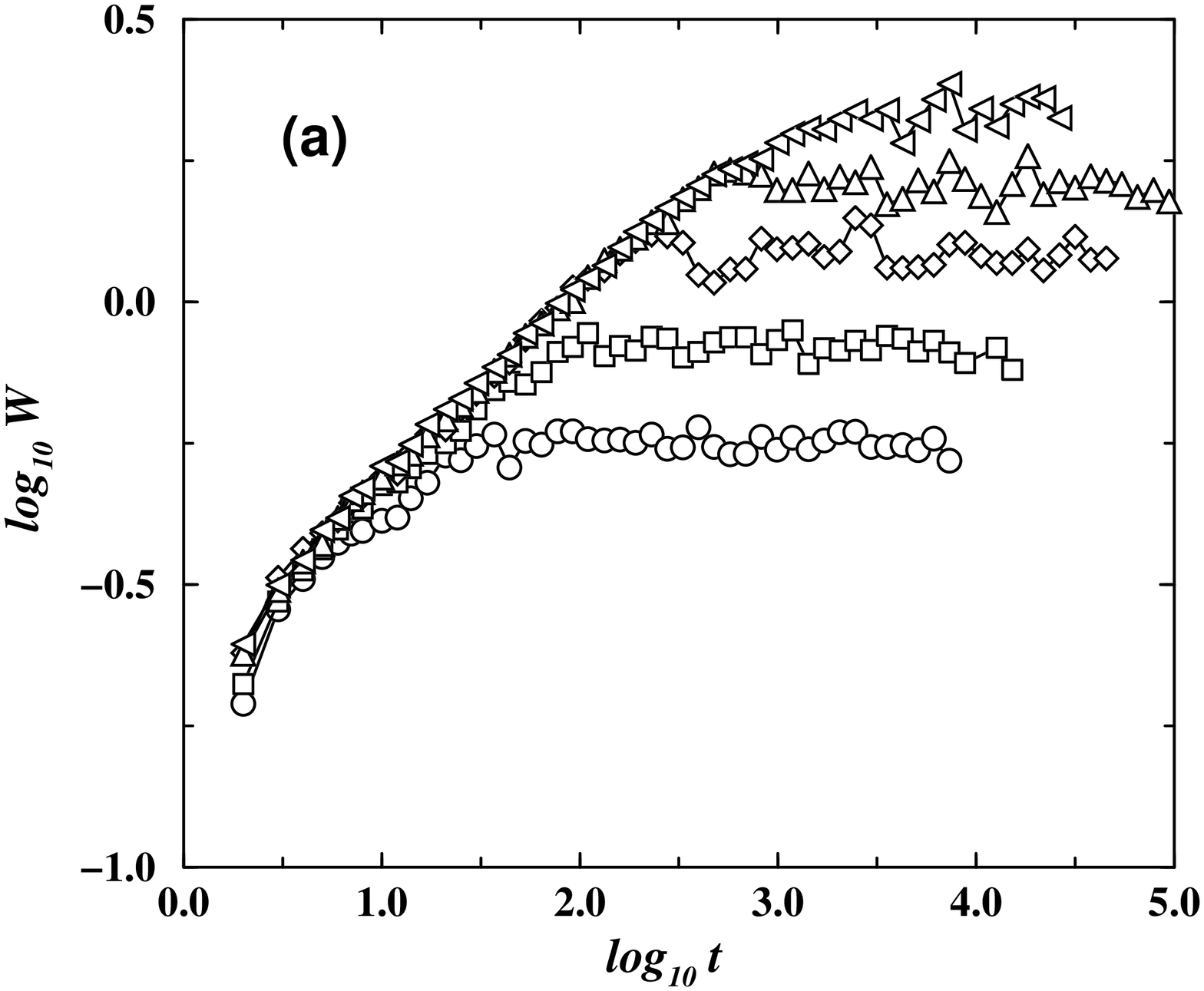,width=3.375in,angle=-90}
\newbox\figb
\setbox\figb=\psfig{figure=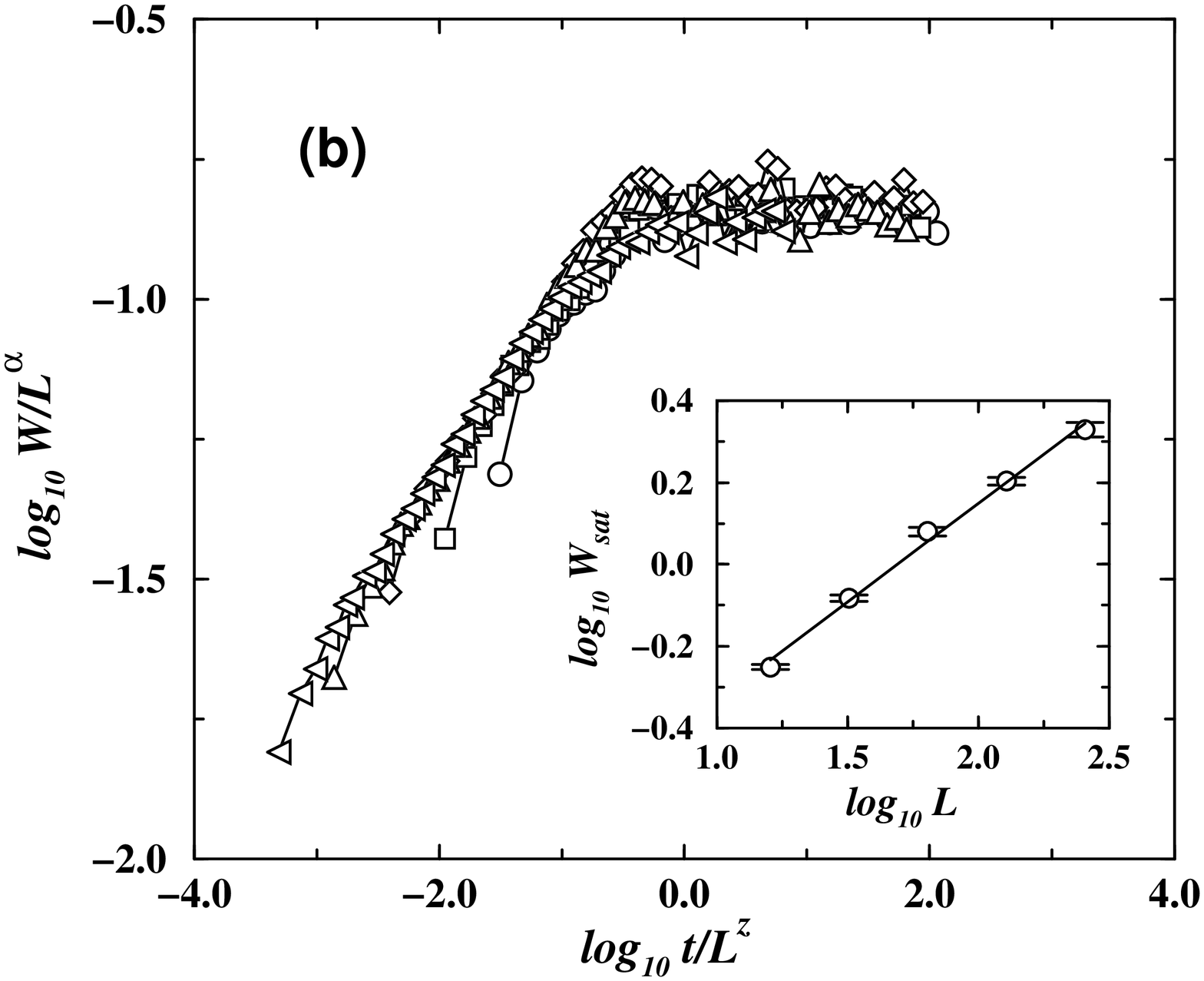,width=3.375in,angle=-90}
\begin{figure*}[htbp]
\begin{center}
\leavevmode
\centerline{\box\figa}
\centerline{\box\figb}
\narrowtext
\caption{ (a) Interface growth and roughening of a binary fluid for
system sizes $L=16 (\circ),32 (\Box),64 (\Diamond),128 (\triangle)$ and
$256 (\lhd)$.  At early times, the width $W \sim t^{\beta}$ where $\beta
= 0.33 \pm 0.02$.  (b) Rescaling to test the scaling hypothesis and the
values of $\alpha$ and $\beta$.  Error bars are approximately equal to
the symbol size.  The inset shows the size scaling of saturation width,
where $\alpha = 0.50 \pm 0.02$, in agreement with the values predicted
in ref.~\protect\cite{FR}.}
\label{Fig1}
\end{center}
\end{figure*}

\noindent line up along the interface between the two fluids.  Since we
are interested in the role that the surfactant plays in the kinetic
roughening process, we start with an initial configuration where the
fluids are separated and we replace red and blue particles with
surfactant particles in a {\it single} row along the interface
(Fig.~\ref{Fig0}(b)).  We define the surfactant concentration $\rho_s
\equiv N_s / L$, where $N_s$ is the number of surfactant particles
placed along the interface.  To obtain $\rho_s > \rho$, we replace red
and blue particles with surfactant particles on {\it multiple} rows
\cite{conc2}.

In Fig.~\ref{Fig2}(a), we plot $W(L,t)$ for all system sizes and $\rho =
\rho_s=0.5$.  We still see a growth regime that crosses 

\newbox\figa
\setbox\figa=\psfig{figure=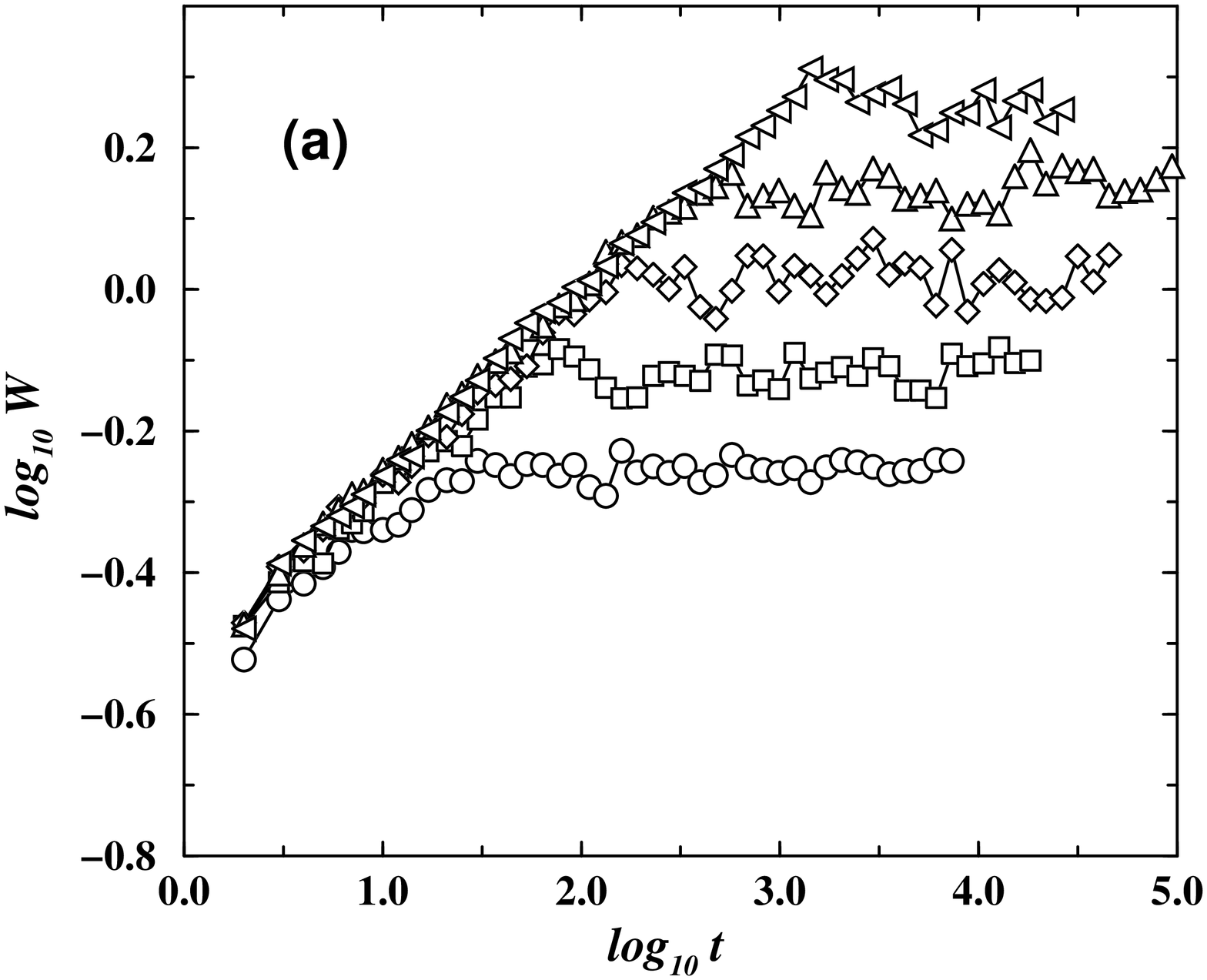,width=3.375in,angle=-90}
\newbox\figb
\setbox\figb=\psfig{figure=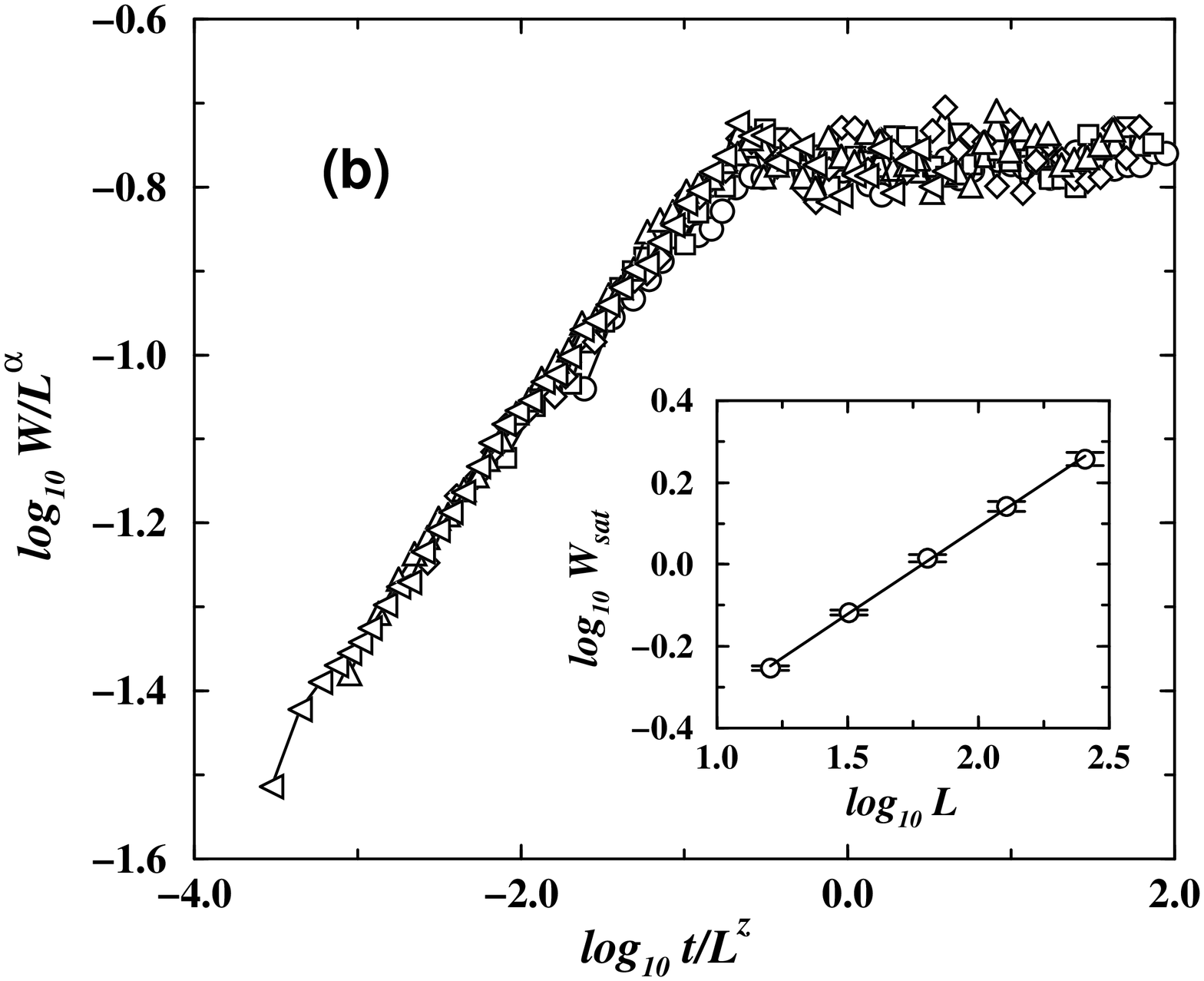,width=3.375in,angle=-90}
\begin{figure*}[htbp]
\begin{center}
\leavevmode
\centerline{\box\figa}
\centerline{\box\figb}
\narrowtext
\caption{ (a) Interface growth and roughening for the three component
model with a surfactant concentration $\rho_s = 0.5$ along the
interface.  System sizes and symbols are the same as
Fig.~\protect\ref{Fig1}.  For $t < t_{\times}$, the growth rate is
slower than for the binary case, and the saturation width is smaller.
We find $\beta = 0.27 \pm 0.02$.  (b) Rescaled plot demonstrating data
collapse.  Error bars are approximately equal to the symbol size.  The
inset shows the roughness exponent $\alpha = 0.43 \pm 0.02$.}
\label{Fig2}
\end{center}
\end{figure*}

\noindent over to a
saturated regime.  Compared to Fig.~\ref{Fig1}, the case without
surfactant, the growth rate of the width is slower and $W_{sat}$ is
smaller.  We find $\beta=0.27 \pm 0.02$ and $\alpha = 0.43 \pm 0.02$.
Fig.~\ref{Fig2}(b) confirms the scaling hypothesis using $z = \alpha /
\beta = 1.59 \pm 0.03$.  These values for $\alpha$ and $\beta$ are not
the accepted values for the roughness and growth exponents of the KPZ
equation in $1+1$ dimensions.  The KPZ exponents also obey the sum rule,
$z+\alpha = 2$, a consequence of Galilean invariance \cite{vfk}.  We
obtain $z + \alpha = 2.02 \pm 0.04$ for this case.  This result is
surprising since lattice-gas models are not Galilean invariant, due to
the fact that the lattice constitutes a preferred Galilean reference
frame \cite{FHP}.  On the other hand, 

\newbox\figa
\setbox\figa=\psfig{figure=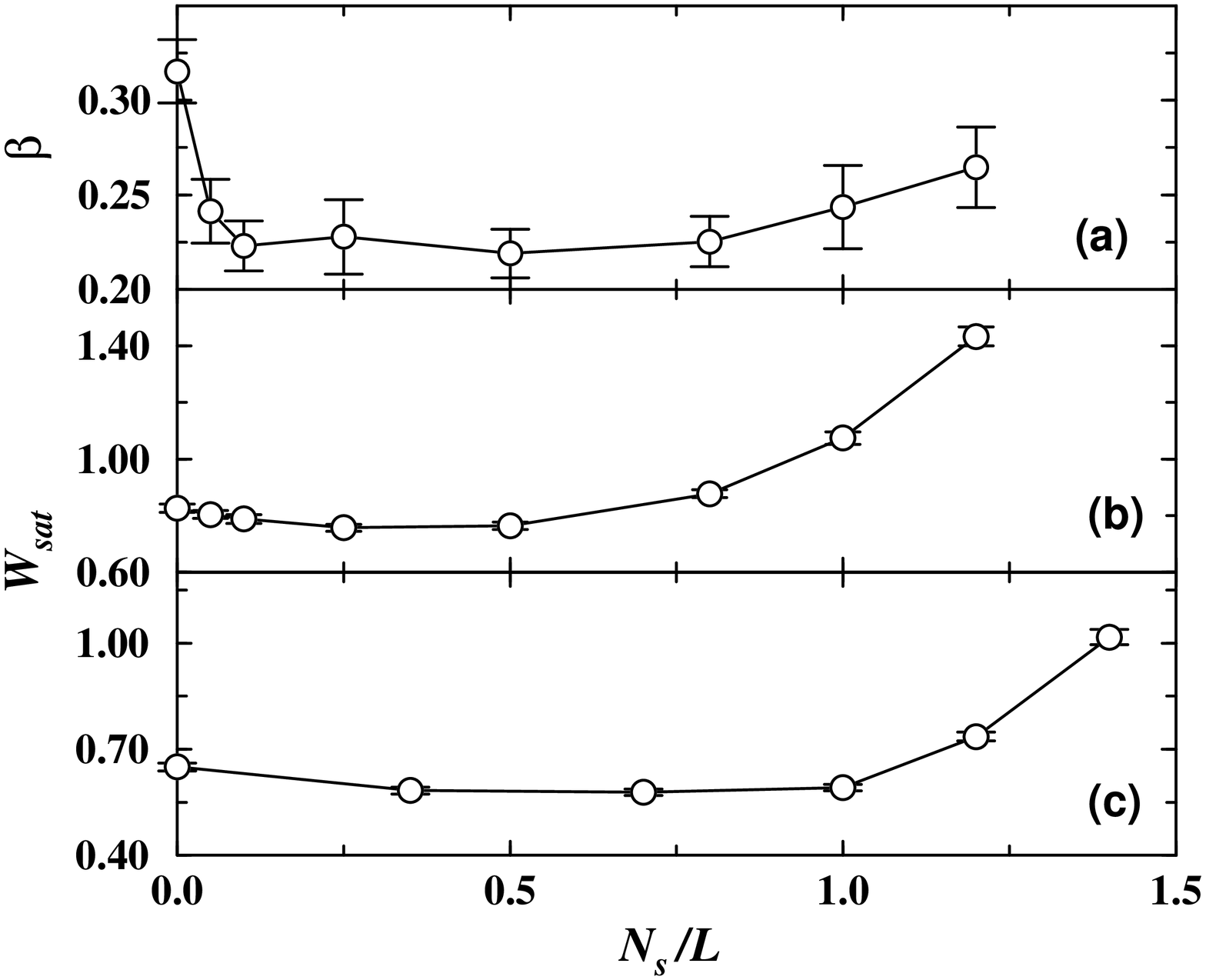,width=3.375in,angle=-90}
\begin{figure*}[htbp]
\begin{center}
\leavevmode
\centerline{\box\figa}
\narrowtext
\caption{ (a) Dependence of the growth exponent $\beta$ on surfactant
concentration $\rho_s \equiv N_s/L$ for a $32 \times 32$ system with $\rho =
0.5$.  For small surfactant concentrations, the growth rate is reduced.
$\beta$ increases for $\rho_s > \rho$ until spontaneous emulsification
occurs.  (b) Dependence of saturation width on surfactant concentration
for $\rho = 0.5$.  For $\rho_s < \rho$ the surfactant reduces the value
of the saturated width.  For $\rho_s > \rho$ the interface becomes
rougher as it approaches spontaneous emulsification.  (c) $W_{sat}$
dependence on $\rho_s$ with a binary fluid concentration $\rho = 0.7$.
We again see that the behavior changes at $\rho_s = \rho$.}
\label{Fig3}
\end{center}
\end{figure*}

\noindent the flow equations of the
lattice-gas model do contain a non-linear inertial term, as does the KPZ
equation.

We next consider the effect of the surfactant concentration on the
growth of the interface and the saturated width.  In Fig.~\ref{Fig3}(a),
we show the dependence of $\beta$ on $\rho_s$ for a lattice of size $32
\times 32$.  We observe a dramatic reduction in the growth exponent upon
the addition of any surfactant to the interface.  Additional surfactant
causes little change in $\beta$ up to $\rho_s=\rho$, where $\beta = 0.22
\pm 0.02$.  For $\rho_s > \rho$, the growth exponent increases but never
returns to the value of the pure case.  For $\rho_s \gtrsim 1.5$, the
interface spontaneously breaks up, giving rise to a micellular
microemulsion phase where the width is no longer well defined (Fig.
1(c)).  The effect of $\rho_s$ on $W_{sat}$ is similar to its effect on
$\beta$.  For $\rho_s < \rho$ we see that the interface width saturates
at a value that decreases for increasing $\rho_s$ (Fig.~\ref{Fig3}(b)).
However, for $\rho_s > \rho$, $W_{sat}$ increases until we reach the
microemulsion phase.

We can understand the behavior of $W_{sat}$ by considering the effect
the surfactant has on the binary fluid interface.  From the Hamiltonian,
we know that the color-dipole energy, ${\bf \sigma} \cdot {\bf E}$,
required for surfactant to be removed from the interface is large
compared to the energy for the surfactant to remain on the interface.
As a result, the surfactant particles are effectively ``anchored'' at
the interface.  For $\rho_s < \rho$, the surfactant reduces the
fluctuations of the interface because it is anchored in position between
red and blue particles.  As a result, $W_{sat}$ is smaller.  For $\rho_s
> \rho$, the color-dipole interaction still forces surfactant particles
to remain on the interface, but now surfactant particles are more likely
to neighbor each other.  Since the dipole-dipole interaction, ${\bf
\sigma} \cdot {\bf P}$, makes it energetically unfavorable for dipole
molecules to align with other dipole molecules at neighboring sites, the
surfactant creates additional surface by roughening the interface to
reduce the number of neighboring surfactant molecules, thereby
increasing $W_{sat}$.  To demonstrate that the turning point in
$W_{sat}$ depends on our choice of $\rho$, we measure $W_{sat}$ for
various $\rho_s$ with a binary fluid concentration $\rho = 0.7$
(Fig.~\ref{Fig3}(c)).  We again see $W_{sat}$ decreasing for $\rho_s \le
\rho$ and increasing for $\rho_s > \rho$.

To investigate the effect of $\rho_s$ on $\alpha$, we consider system
sizes $L=16, 32, 64$ and $128$ for $\rho_s = 0.8$ and $1.0$.  At $\rho_s
= 0.8$ we find $\beta = 0.27 \pm 0.02$ and $\alpha = 0.36 \pm 0.02$,
while for $\rho_s = 1.0$ we obtain $\beta = 0.30 \pm 0.02$ and $\alpha =
0.23 \pm 0.02$.  Using the exponents for these concentrations, we do not
find $z+\alpha=2$.  In Table I we summarize the behavior of the
exponents $\alpha$, $\beta$, $z$, and $z+ \alpha$ as a function of the
surfactant concentration.  We notice that $\alpha$ decreases as we
increase the surfactant concentration.

In summary, we have found that a three-component lattice gas can be used
to test the effect of surfactant of the scaling properties of
hydrodynamic interfacial growth and roughening.  In particular, we have
found that surfactant alters the growth and roughness exponents, with
$\alpha$ decreasing and $\beta$ increasing as we increase $\rho_s$ above
the concentration of the binary fluid.  These results may assist in
understanding the stability limits of interfacial growth in the presence
of a surfactant and the conditions for the formation and breakup of
microemulsion droplets.

\begin{center}
\begin{minipage}{3.25in}
\begin{table}
\caption{Roughness, growth and dynamic exponents for various surfactant
concentrations measured from the simulations.  Also included are the
number of samples that have been averaged.}
\medskip
\begin{tabular}{cc|ccccc|c}
$\rho_s$ & $$ & $\alpha$ & $\beta$ & $z$ & $z+\alpha$ & $$ &	Samples\\
\tableline
$0.0$ & $$ & $0.50 \pm 0.02$ & $0.33 \pm 0.02$ & $1.50 \pm 0.03$ & $2.00 \pm 0.04$ & $$ & $170$ \\
$0.5$ & $$ & $0.43 \pm 0.02$ & $0.27 \pm 0.02$ & $1.59 \pm 0.03$ & $2.02 \pm 0.04$ & $$ & $170$ \\
$0.8$ & $$ & $0.36 \pm 0.02$ & $0.27 \pm 0.02$ & $1.33 \pm 0.03$ & $1.69 \pm 0.04$ & $$ & $150$ \\
$1.0$ & $$ & $0.23 \pm 0.02$ & $0.30 \pm 0.02$ & $0.77 \pm 0.03$ & $1.00 \pm 0.04$ & $$ & $150$ \\
\end{tabular}
\label{table1}
\end{table}
\end{minipage}
\end{center}

We thank A.N.~Emerton for the use of his lattice-gas code,
L.A.N.~Amaral, S.V.~Buldyrev, P.V.~Coveney, S.~Havlin, and D.H.~Rothman
for helpful discussions, and NSF for financial support.  FWS and STH are
supported by NSF and NIH predoctoral fellowships, respectively, while
BMB is supported in part by Phillips Laboratories and by the United
States Air Force Office of Scientific Research under grant number
F49620-95-1-0285.  Simulations were carried out using the Power
ChallengeArray at the Boston University Center for Computational
Science, and the IBM SP2 at the Maui High Performance Computing Center.

\end{multicols}

\begin{references}
\bibitem{coupling} The relative strength of each term in the Hamiltonian
can be controlled by coupling constants.  Using the nomenclature of
ref.~\cite{BCE}, we have chosen $\alpha = 1.0, \mu = 0.05, \epsilon =
8.0$, and $\zeta = 0.5$.  Here $\alpha$ is not to be confused with the
roughness exponent.
\bibitem{conc} This is equivalent to 3.5 particles per site on average.
This choice is not crucial for a binary fluid provided $\rho$ is above
the critical concentration needed for phase separation.
\bibitem{conc2} For $\rho_s \le \rho$, we replace the red and blue
particles with surfactant particles on only one row of the interface.
For $\rho < \rho_s \le 2 \rho$, we use two rows.  Similarly, we
use three rows for $2 \rho < \rho_s \le 3 \rho$.
\end{references}
\end{document}